\documentclass[english,twocolumn, pra]{revtex4-1}
\usepackage[T1]{fontenc}
\usepackage[latin9]{inputenc}
\usepackage{amsmath}
\usepackage{graphicx}
\usepackage{esint}
\usepackage{babel}
\begin{document}

\title{Stochastic dressed wavefunction: a numerically exact solver for bosonic
impurity model dynamics within wide time interval}

\author{Evgeny A. Polyakov$^{1}$, Alexey N. Rubtsov$^{1,2}$}

\address{$^{1}$Russian Quantum Center, 100 Nonaya St., Skolkovo, Moscow 143025,
Russia}

\address{$^{2}$Department of Physics, Lomonosov Moskow State University,
Leninskie gory 1, 119991 Moscow, Russia}
\begin{abstract}
In the dynamics of driven impurity models, there is a fundamental
asymmetry between the processes of emission and absorption of environment
excitations: most of the emitted excitations are rapidly and irreversibly
scattered away, and only a small amount of them is reabsorbed back.
We propose to use a stochastic simulation of the irreversible quantum
emission processes in real-time dynamics, while taking into account
the reabsorbed virtual excitations by the bath discretization. The
resulting method delivers a fast convergence with respect to the number
of bath sites, on a wide time interval, without the sign problem.
\end{abstract}
\maketitle

\section{INTRODUCTION}

The quantum impurity model has always been the cornerstone in condensed
matter and quantum optics. Introduced in order to describe the interaction
of magnetic impurities with a metallic host \cite{Hewson1993}, this
model is used to describe low-temperature properties of single-electron
solid-state devices \cite{Kastner1992,Goldhaber-Gordon1998}, tunelling
spectroscopy experiments \cite{Manoharan2000,Agam2000}, mobility
of defects \cite{Zimmerman1991,Golding1992} and of interstitials
\cite{Kondo1984,Wipf1984,Grabert1997} in solids. In the fields of
quantum optics and quantum information processing, driven impurity
model in a bosonic environment is often called an open quantum system.
It is used to describe the two-level atoms in optical fibers \cite{LeClair1997},
Cooper pair boxes coupled to an electromagnetic environment \cite{LeHur2012,Goldstein2013,Peropadre2013,Snyman2015,Bera2016}
and solid-state qubits \cite{Daniel1998}. In physical chemistry the
quantum imputiry model is employed in theoretical analysis of the
electron transfer processes between donor and acceptor molecules \cite{Marcus1985,Tornow2008}. 

Lately there was a revisited interest to a numerically exact solvers
of the impurity model in a situation when its coupling to the bath
is not small. Initially it was connected to the development of dynamical
mean-field theory (DMFT) calculations \cite{Pruschke1995,Georges1996,Freericks2006,Aoki2014}.
Within DMFT and its cluster extensions \cite{Maier2005}, lattice
models for strongly correlated fermions are mapped onto quantum impurity
problems which are embedded into environment whose spectral properties
are determined self-consistently. These equilibrium fermionic problems
required solvers of the Anderson impurity working at imaginary-time
Matsubara domain. A continuous time quantum Monte Carlo (CT-QMC) family
of algorithms \cite{Gull2011} was constructed to deliver results
which are free of any systematic errors and obey a reasonably small
stochastic noise. Experiments with ultracold atomic systems driven
the efforts to construct impurity solvers for real time dynamics away
from equilibrium \cite{Strand2015,Panas2015,Panas2017}. In this case,
both fermionic and bosonic systems are of importance. For bosonic
ones, an additional interest is related with cavity-QED and similar
problems, where one deals with a (driven) two level system strongly
coupled to phonons.

A generic problem about the real-time impurity solvers is that the
computational complexity scales exponentially with the increasing
time argument. The physical origin of the problem is that as the time
passes, the quantum impurity scatters environmental excitations with
a (roughly) constant rate. As a consequence, the number of mutually
entangled excitations increases at least linearly with time, and thus
the dimension of the relevant entangled subspace of the total Hilbert
space increases exponentially. In different simulation techniques,
this basic issue manifests itself in distinct ways. In the basis truncation
methods, we need to include exponentially large number of basis elements
as the simulation time is increased. The density matrix renormalization
group (DMRG) \cite{Wong2008} and numerical renormalization group
(NRG) \cite{Vojta2012} methods also entail the truncation of Hilbert
space, and this limits the range of parameters where results of sufficient
accuracy can be obtained. In the quantum Monte Carlo (QMC) simulation
techniques \cite{Egger2000,Maier2005,Needs2010,LeBlanc2015}, the
complexity comes out as the sign problem due to the oscillating phase
factors of trajectories (diagramms). The quasi-adiabatic path integral
(QUAPI) approach \cite{Makarov1994,Makri1995,Makri1995a,Makri1996}
has convergence problems at low temperatures and when the environment
memory is long \cite{Chen2017a,Segal2010}. The hierarchical equations-of-motion
(HEOM) method \cite{Taminura1989,Ishizaki2009,Strumpfer2012} employs
a Matsubara expansion for the bath density matrix. HEOM is accurate
at high temperatures and for near-Debye spectral densities \cite{Chen2017a},
but displays exponential complexity as we move outside these case.
The multi-layer multi-configuration time-dependent Hartree (ML-MCTDH)
approach \cite{Thoss2001,Wang2001,Wang2003} has problems in the strongly
correlated regimes \cite{Chen2017a,Wang2013}. Probably the most promising
of existing real-time solvers is so-called inchworm QMC algorithm
\cite{Cohen2015,Chen2017,Chen2017a}, in which the Keldysh contour
is split to a number of intervals and the diagrams are hierarchically
summed up on them. The method alleviates the sign problem to a large
degree, but is of high technical complexity and suffers from a fast
grow of memory requirements as time scale increases. 

In this paper we propose a technically simple and physically transparent
real-time bosonic impurity solver, which is free from a sign problem
and does not show signs of an exponential slow down for a number of
benchmark problems. In our approach, virtual bath excitations and
really emitted bosons are treated in different ways: real (observable)
excitations are accounted for within a sign-free QMC (stochastic)
procedure, whereas virtual ones are described by the ED treatment.
With an increase of time argument, only the number of real excitation
grows, that allows to escape an exponential increase of the Fock space
for the ED part.

In section \ref{sec:DESCRIPTION-OF-THE} we introduce the general
impurity model in bosonic bath. Then in section \ref{subsec:Influence-functional-and}
we recall the Keldysh contour path integral formalism and discuss
the physical interpretation of influence functional which describes
the effect of the bath on the impurity. Using the acquired intuition,
in section \ref{subsec:An-idea-of} we identify the major factors
leading to the exponential complexity of real-time quantum simulation.
We formulate the algorithm enabling us to alleviate these factors
in \ref{subsec:The-stochastic-dressed} and \ref{subsec:Approximate-solution-of}.
The results of test calculations for the spin-boson model are presented
in section \ref{sec:RESULTS}. Finally, we conclude in section \ref{sec:CONCULSION}.

\section{\label{sec:DESCRIPTION-OF-THE}DESCRIPTION OF THE METHOD}

In this section, we present our approach to the simulation of open
quantum system dynamics. We consider the following impurity system
Hamiltonian
\begin{equation}
\widehat{H}=\widehat{H}_{\textrm{i}}+\widehat{V}+\widehat{H}_{\textrm{b}},
\end{equation}
where $\widehat{H}_{\textrm{i}}$ and $\widehat{H}_{\textrm{b}}$
are the Hamiltonians of the impurity and of the bath, respectively,
and $\widehat{V}$ is a system-environment interaction. The environment
is supposed to have a quadratic Hamiltonian 
\begin{equation}
\widehat{H}_{\textrm{b}}=\intop_{-\infty}^{+\infty}d\omega\omega\widehat{b}^{\dagger}\left(\omega\right)\widehat{b}\left(\omega\right),
\end{equation}
with the bilinear interaction 
\begin{equation}
\widehat{V}=\widehat{s}\widehat{b}^{\dagger}+\widehat{s}^{\dagger}\widehat{b},
\end{equation}
where $\widehat{s}$ is a certain impurity operator, and $\widehat{b}$
is the bath degree of freedom
\begin{equation}
\widehat{b}=\intop_{-\infty}^{+\infty}d\omega c\left(\omega\right)\widehat{b}\left(\omega\right).
\end{equation}
In our representation, the frequency dependence of the density-of-states
is transfered to the coupling coefficient $c\left(\omega\right)$.
We are interested in the calculation of the time-dependent impurity
observable mean values:
\begin{equation}
\left\langle \widehat{O}\left(t\right)\right\rangle =\textrm{Tr}\left\{ \widehat{O}e^{-it\widehat{H}}\widehat{\rho}\left(0\right)e^{it\widehat{H}}\right\} .\label{eq:operator_average}
\end{equation}
Here $\widehat{\rho}_{0}$ is the initial state of the total system.
The trace operation $\textrm{Tr}\left\{ \cdot\right\} $ is taken
over all states of the full system. Let us make the conventional assumption
that the initial state is factorized, 
\begin{equation}
\widehat{\rho}\left(0\right)=\left|\psi_{\textrm{i}}\left(0\right)\right\rangle \left\langle \psi_{\textrm{i}}\left(0\right)\right|\otimes\widehat{\rho}_{\textrm{b}}\left(0\right),\label{eq:factorized_initial_state}
\end{equation}
where $\psi_{\textrm{i}}\left(0\right)$ is arbitrary state in the
impurity's Hilbert space, and $\widehat{\rho}_{\textrm{b}}\left(0\right)$
is assumed to be a Gaussian bath state with certain mode occupations
\begin{equation}
n\left(\omega\right)=\textrm{Tr}_{\textrm{b}}\left\{ \widehat{b}^{\dagger}\left(\omega\right)\widehat{b}\left(\omega\right)\widehat{\rho}_{\textrm{b}}\left(0\right)\right\} .
\end{equation}
Here, $\textrm{Tr}_{\textrm{b}}\left\{ \cdot\right\} $ denotes the
trace over the bath degrees of freedom. 

\subsection{\label{subsec:Influence-functional-and}Influence functional and
its physical interpretation}

In order to understand the physical structure of the driven impurity
problem, it will be helpfull to express the observable mean value
Eq. (\ref{eq:operator_average}) in terms of the Keldysh functional
integral \cite{Kamenev2011}, and employ the notion of influence functional
of the environment \cite{Feynman1963,Weiss2012}.

\begin{figure}
\includegraphics{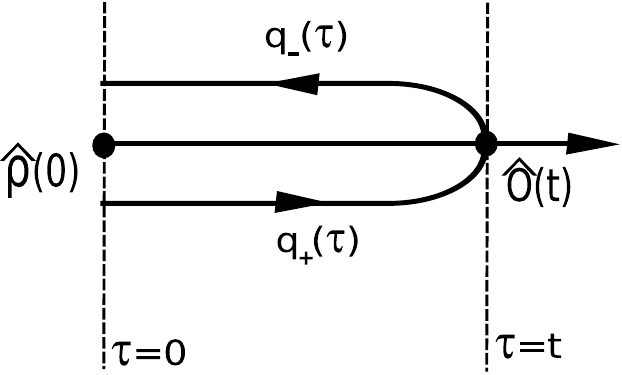}

\caption{\label{fig:keldysh_contour}The Keldysh contour represents the closed-time
evolution of the impurity, starting from the time $\tau=0$ in the
initial state $\widehat{\rho}\left(0\right)$, forward in time (the
lower branch whose quantities are labeled by subscript ``+'') up
to the time $\tau=t$. Here the observable $\widehat{O}$ is inserted.
Then the system evolves backwards in time (the upper branch whose
quantities are labeled by subscript ``-'') up to the initial time
$\tau=0$. }
\end{figure}

Let us consider the following general real-time quantum problem

\begin{equation}
\left\langle \widehat{O}\left(t\right)\right\rangle =\textrm{Tr}\left\{ \widehat{\rho}_{\textrm{out}}\left(t\right)\widehat{O}e^{-it\widehat{H}}\widehat{\rho}_{\textrm{in}}\left(0\right)e^{it\widehat{H}}\right\} ,\label{eq:in-out-problem}
\end{equation}
where $\widehat{O}$ is the impurity observable. With the choice 
\begin{equation}
\widehat{\rho}_{\textrm{in}}\left(0\right)=\left|\psi_{\textrm{i}}\left(0\right)\right\rangle \left\langle \psi_{\textrm{i}}\left(0\right)\right|\otimes\widehat{\rho}_{\textrm{b}}\left(0\right),\label{eq:standard_input_state}
\end{equation}
\begin{equation}
\widehat{\rho}_{\textrm{out}}\left(t\right)=\widehat{1}_{\textrm{i}}\otimes\widehat{1}_{\textrm{b}},\label{eq:standard_output_state}
\end{equation}
we obtain the problem Eq. (\ref{eq:operator_average}) we are aiming
at. However, in order to derive our method, we also will need to consider
an auxiliary problem with 
\begin{equation}
\widehat{\rho}_{\textrm{in}}\left(0\right)=\left|\psi_{\textrm{i}}\left(0\right)\right\rangle \left\langle \psi_{\textrm{i}}\left(0\right)\right|\otimes\left|0_{\textrm{b}}\right\rangle \left\langle 0_{\textrm{b}}\right|,\label{eq:vacuum_input_state}
\end{equation}
\begin{equation}
\widehat{\rho}_{\textrm{out}}\left(t\right)=\widehat{1}_{\textrm{i}}\otimes\left|0_{\textrm{b}}\right\rangle \left\langle 0_{\textrm{b}}\right|,\label{eq:vacuum_output_state}
\end{equation}
 The Keldysh contour technique allows one to map the quantum problem
(\ref{eq:in-out-problem}) onto the functional integral \cite{Kamenev2011}
over the configurational space of the system, Fig . \ref{fig:keldysh_contour}:
\begin{multline}
\left\langle \widehat{O}\left(t\right)\right\rangle =\int D\left[q_{+},q_{-}\right]\exp\left(iS_{\textrm{i}}\left[q_{+},q_{-}\right]+I\left[q_{+},q_{-}\right]\right)\\
\times O\left(q_{+}\left(t\right),q_{-}\left(t\right)\right).\label{eq:open_system_keldysh_integral}
\end{multline}

Here, $q_{+}\left(\tau\right)$, $q_{-}\left(\tau\right)$ are the
configurational variables of the system on the forward and on the
backward banches of the contour. $S_{\textrm{i}}\left[q_{+},q_{-}\right]$
is the action functional of the impurity. $I\left[q_{+},q_{-}\right]$
is the influence functional of the bath \cite{Feynman1963,Weiss2012},
\begin{multline}
I\left[q_{+},q_{-}\right]\\
=-\iintop_{0}^{t}d\tau d\tau^{\prime}\left[\begin{array}{c}
-s_{+}^{*}\left(\tau\right)\\
s_{-}^{*}\left(\tau\right)
\end{array}\right]^{T}\mathbf{K}\left(\tau-\tau^{\prime}\right)\left[\begin{array}{c}
-s_{+}\left(\tau^{\prime}\right)\\
s_{-}\left(\tau^{\prime}\right)
\end{array}\right].
\end{multline}
Here 
\begin{equation}
s_{\pm}\left(\tau\right)=s\left(q_{\pm}\left(\tau\right)\right)
\end{equation}
is a forward/backward-branch path integral representation of the impurity
operator $\widehat{s}$. The 2-by-2 matrix $\mathbf{K}\left(\tau-\tau^{\prime}\right)$
is the Keldysh correlation function of the bath, 
\begin{equation}
K_{\pm,\pm}\left(\tau-\tau^{\prime}\right)=\textrm{Tr}\left\{ \mathcal{C}\widehat{\rho}_{\textrm{out}}\left(t\right)\widehat{b}\left(\tau_{\pm}\right)\widehat{b}^{\dagger}\left(\tau_{\pm}^{\prime}\right)\widehat{\rho}_{\textrm{in}}\left(0\right)\right\} .
\end{equation}
The contour ordering $\mathcal{C}$ places the operators (as functions
of contour parameter) in the descending order, from left to right.
The contour order is defined as 
\begin{equation}
t\succ_{\mathcal{C}}\tau_{+}\succ_{\mathcal{C}}0,
\end{equation}
\begin{equation}
\tau_{+}\succ_{\mathcal{C}}\tau_{+}^{\prime}\,\,\,\textrm{if}\,\,\,\tau_{+}>\tau_{+}^{\prime},
\end{equation}
\begin{equation}
\tau_{-}\succ_{\mathcal{C}}t,
\end{equation}
\begin{equation}
\tau_{-}\succ_{\mathcal{C}}\tau_{-}^{\prime}\,\,\,\textrm{if}\,\,\,\tau_{-}<\tau_{-}^{\prime},
\end{equation}
\begin{equation}
\tau_{-}\succ_{\mathcal{C}}\tau_{+}^{\prime}.
\end{equation}
For the usual Keldysh contour with factorized inital condition, Eqs.
(\ref{eq:standard_input_state}) - (\ref{eq:standard_output_state}),
we have the following Keldysh correlation function:
\begin{multline}
\mathbf{K}\left(\tau-\tau^{\prime}\right)=\mathbf{K}_{\textrm{virt}}\left(\tau-\tau^{\prime}\right)\\
+\mathbf{K}_{\textrm{emit}}\left(\tau-\tau^{\prime}\right)+\mathbf{K}_{\textrm{exc}}\left(\tau-\tau^{\prime}\right).
\end{multline}
Each of these terms has distinct physical interpretation, as will
become evident below. The first term 
\begin{multline}
\mathbf{K}_{\textrm{virt}}\left(\tau-\tau^{\prime}\right)\\
=\left[\begin{array}{cc}
\theta\left(\tau-\tau^{\prime}\right) & 0\\
0 & \theta\left(\tau^{\prime}-\tau\right)
\end{array}\right]M\left(\tau-\tau^{\prime}\right)
\end{multline}
describes the effect of the virtual (unobservable) bath excitations.
The second term 
\begin{equation}
\mathbf{K}_{\textrm{emit}}\left(\tau-\tau^{\prime}\right)=\left[\begin{array}{cc}
0 & 0\\
1 & 0
\end{array}\right]M\left(\tau-\tau^{\prime}\right)
\end{equation}
describes the irreversible spontaneous emission of observable excitations,
where the bath memory function is
\begin{equation}
M\left(\tau-\tau^{\prime}\right)=\intop_{-\infty}^{+\infty}d\omega\left|c\left(\omega\right)\right|^{2}e^{-i\omega\left(\tau-\tau^{\prime}\right)}.
\end{equation}
 The last term 
\begin{equation}
\mathbf{K}_{\textrm{exc}}\left(\tau-\tau^{\prime}\right)=\left[\begin{array}{cc}
1 & 1\\
1 & 1
\end{array}\right]M_{\textrm{exc}}\left(\tau-\tau^{\prime}\right)
\end{equation}
represents the effects of the quantum excitations of bath due to a
finite initial occupation $n\left(\omega\right)$ of the frequency
modes. Here the excitation noise memory function is
\begin{equation}
M_{\textrm{exc}}\left(t-t^{\prime}\right)=\intop_{-\infty}^{+\infty}d\omega n\left(\omega\right)\left|c\left(\omega\right)\right|^{2}e^{-i\omega\left(t-t^{\prime}\right)}.
\end{equation}
 In most practical situations, $n\left(\omega\right)$ is the finite
temperature Bose-Einstein distribution, 
\begin{equation}
n\left(\omega\right)=\frac{1}{e^{\beta\left(\omega-\mu\right)}-1}.
\end{equation}
The physical interpretation of $\mathbf{K}_{\textrm{exc}}\left(\tau-\tau^{\prime}\right)$
is evident from the fact that this term vanishes when the bath is
initially in the vacuum state.

\begin{figure}
\includegraphics[scale=1.3]{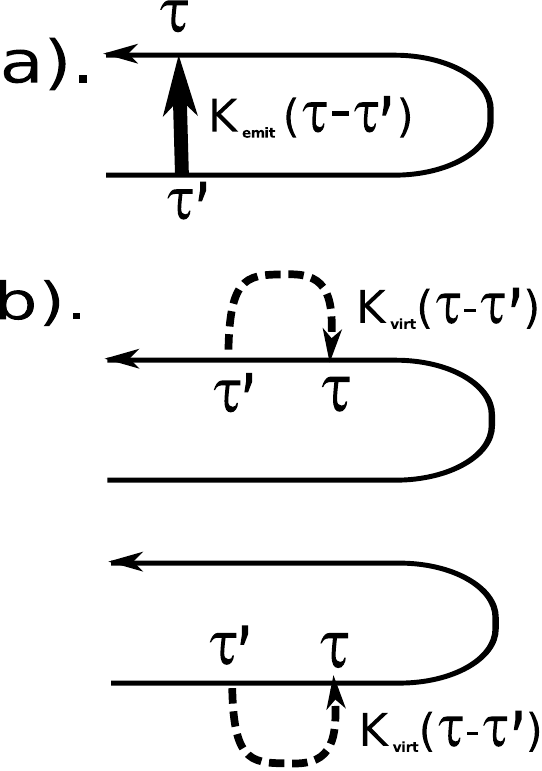}

\caption{\label{fig:types_of_propagators}a). Each factor $\mathbf{K}_{\textrm{emit}}\left(\tau-\tau^{\prime}\right)$
in the perturbation expansion is represented by a bold line crossing
one branch of Keldysh contour at a time point $\tau$ and the other
branch at a time point $\tau^{\prime}$. b) Each factor $-\mathbf{K}_{\textrm{virt}}\left(\tau-\tau^{\prime}\right)$
in the perturbation expansion is represented by a dashed line crossing
the same branch at two time points $\tau$ and $\tau^{\prime}$. }

\end{figure}

In order to illustrate the physical meaning of $\mathbf{K}_{\textrm{virt}}\left(\tau-\tau^{\prime}\right)$
and $\mathbf{K}_{\textrm{emit}}\left(\tau-\tau^{\prime}\right)$,
let us assume that the bath is in vacuum state (there is no $\mathbf{K}_{\textrm{exc}}\left(\tau-\tau^{\prime}\right))$.
We perform the perturbative expansion of the average Eq. (\ref{eq:open_system_keldysh_integral})
with respect to $\mathbf{K}_{\textrm{virt}}\left(\tau-\tau^{\prime}\right)$
and $\mathbf{K}_{\textrm{emit}}\left(\tau-\tau^{\prime}\right)$.
This expansion is represented by a series of diagrams, where each
factor $\mathbf{K}_{\textrm{emit}}\left(\tau-\tau^{\prime}\right)$
is represented by a bold line crossing different branches, and each
factor $-\mathbf{K}_{\textrm{virt}}\left(\tau-\tau^{\prime}\right)$
is represented by a dashed line crossing the same branch, Fig. \ref{fig:types_of_propagators}. 

\begin{figure}
\includegraphics[scale=0.5]{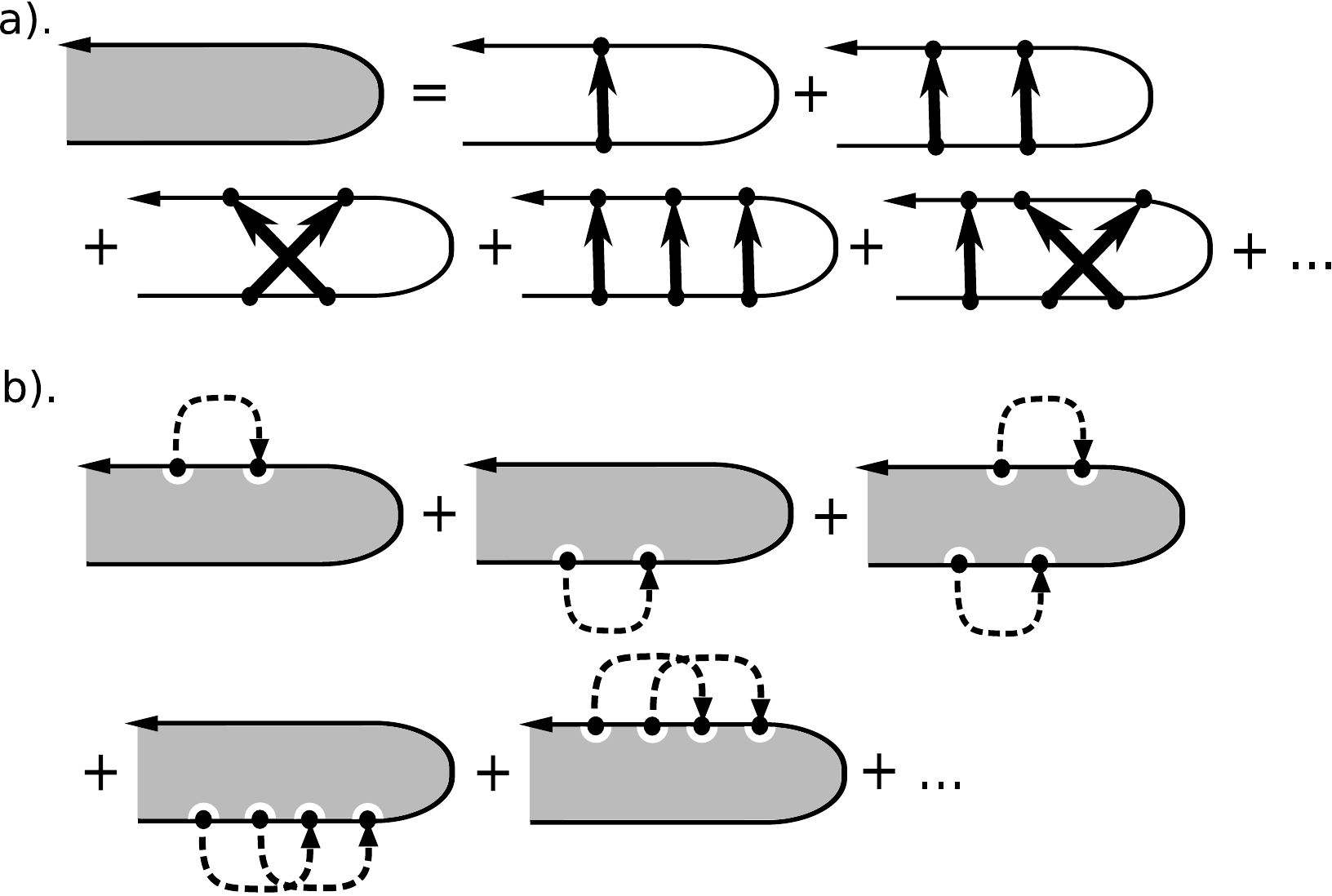}

\caption{\label{fig:diagrammatic_expansion} We classify the diagrams into
the following two classes. a). The first class, which we denote by
a filled Keldysh contour, contains the sum of all the diagrams in
which only the cross-branch lines are present. b). The second class
contain the diagrams with at least one virtual intra-branch line. }
\end{figure}
The whole perturbation expansion consists of the diagrams obtained
by all the posstible insertions of bold and dashed lines, at arbitrary
time points. Then, we observe the following. The time moment of measurement
(where the impurity observable $\widehat{O}$ is placed) is the turning
point of the Keldysh contour, Fig \ref{fig:keldysh_contour}. Therefore,
all the cross-branch lines (with factors $\mathbf{K}_{\textrm{emit}}\left(\tau-\tau^{\prime}\right)$)
correspond to the excitations which exist at the measurement time,
and make a contribution to it, i.e. they are observable, Fig. \ref{fig:types_of_propagators},
a). Whereas all the intrabranch lines (with factors $-\mathbf{K}_{\textrm{virt}}\left(\tau-\tau^{\prime}\right)$)
represent the excitations which are created and annihilated before
the measurement time moment, i.e. they represent the unobservable
virtual excitations, Fig. \ref{fig:types_of_propagators}, b). According
to the aforementioned observation, we divide all the diagrams into
the two classes, Fig. \ref{fig:diagrammatic_expansion}. The first
class, containing the diagramms with only the cross-branch lines,
Fig. \ref{fig:diagrammatic_expansion}, a)., we call the ``cross-branch
diagrams''. They describe the effect of (unread) measurement at time
$t$ of the irreversibly emitted bath-excitation quantum field. The
second class of diagrams, containing at least one virtual intrabranch
line, Fig. \ref{fig:diagrammatic_expansion} b)., which we call the
``intra-branch diagrams'', describe the dynamical effect of the
unobservable cloud of virtual excitations, which always surround any
impurity system.

For the Keldysh contour in which the bath evolves from vacuum to vacuum,
Eqs. (\ref{eq:vacuum_input_state}) and (\ref{eq:vacuum_output_state}),
the Keldysh correlation function in the influence functional Eq. (\ref{eq:open_system_keldysh_integral})
consists of only the virtual part, $\mathbf{K}\left(\tau-\tau^{\prime}\right)=\mathbf{K}_{\textrm{virt}}\left(\tau-\tau^{\prime}\right),$
which again supports our physical interpretation: if there is no emitted
field, and no bath excitations, the virtual excitations still present.

\subsection{\label{subsec:An-idea-of}An idea of how to eliminate the complexity
of real-time simulation}

\begin{figure}
\includegraphics[scale=0.5]{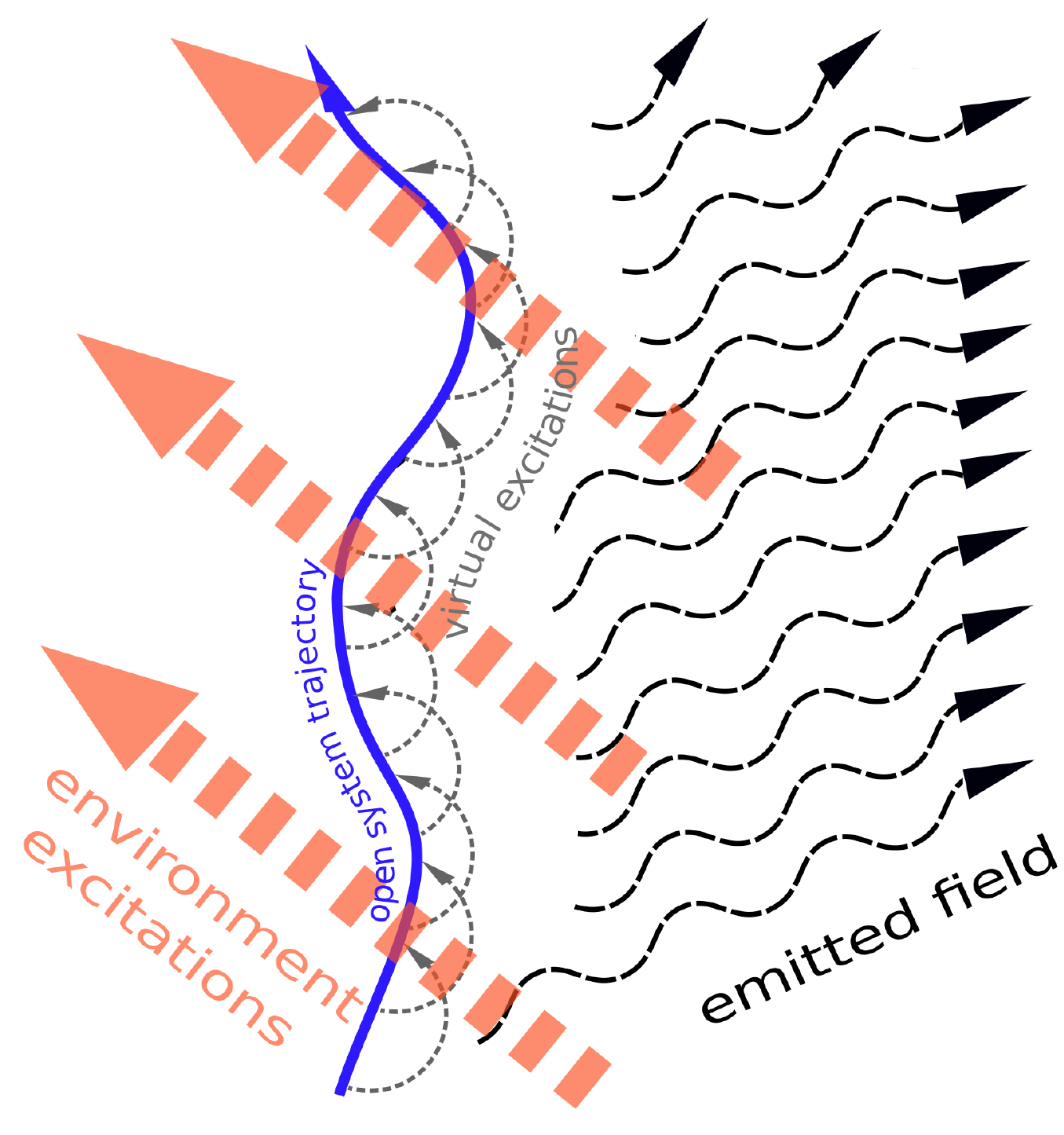}

\caption{\label{fig:oqs_world_line}Suppose that we couple our impurity to
the bath at the time moment $t=0$. Then, the following three processes
start to develop. First, the impurity begin to scatter and to entangle
to the bath excitations. Second, the driven impurity begins to emit
excitations, which also remain entangled to the impurity. Evidently,
the number of excitations involved into these processes will grow
without bound as the time passes. However the process of the third
kind, the emission and the ultimate absorption of virtual excitations,
is expected to saturate on a certain level, so that only a limited
amount of virtual excitations is present.}

\end{figure}

Suppose we have an impurity, and we want to compute its real-time
evolution. The source of the complexity of this problem lies in the
fact that the impurity becomes entangled to the bath excitations,
and as the time goes on, the number of entangled excitations grows
(in most cases) asymptotically linearly with time. As a consequence,
the dimension of the entangled Hilbert subspace grows combinatorially
(exponentially) with time. In the previous section, we have identified
the three parts of the influence functional, which correspond to the
three types of processes: the virtual processes, the irreversible
emission, and the excitations by the bath. Let us analyze the contribution
of each of these parts to the complexity of real-time simulation,
Fig. \ref{fig:oqs_world_line}. The last two processes, the irreversibe
emission and the excitations by the bath, lead to the growth of the
number of entangled excitations. However, the first process, the creation/annihilation
of virtual excitations, is expected to reach a stationary number of
excitations, so that this is not the factor of complexity. 

Then, were it possible to simulate efficiently and in a numerically
exact way the emission and the excitation processes, the complexity
of the real-time simulation would be greatly reduced. Luckily, we
have found at least one way of doing it: the stochastic wavefunction
method \cite{Diosi1997,Diosi1998,Strunz2001,Hartman2017}. 

\subsection{\label{subsec:The-stochastic-dressed}The stochastic dressed wavefunction
method}

In the spirit of stochastic wavefunction method \cite{Diosi1997,Diosi1998,Strunz2001,Hartman2017},
we stochastically unavel the emission and excitation parts of the
influence functional by applying the Hubbard-Stratonovich transform.
Denoting $\boldsymbol{s}\left(\tau\right)=\left[-s_{+}\left(\tau\right),\,s_{-}\left(\tau\right)\right]^{T}$,
we have for the emissive part: 
\begin{multline}
e^{-\iintop_{0}^{t}d\tau d\tau^{\prime}\boldsymbol{s}^{\dagger}\left(\tau\right)\mathbf{K}_{\textrm{emit}}\left(\tau-\tau^{\prime}\right)\boldsymbol{s}\left(\tau^{\prime}\right)}\\
=e^{\iintop_{0}^{t}d\tau d\tau^{\prime}s_{-}^{*}\left(\tau\right)M\left(\tau-\tau^{\prime}\right)s_{+}\left(\tau^{\prime}\right)}\\
=\int D\left[\xi\right]e^{-\intop_{-\infty}^{+\infty}d\omega\left|\xi\left(\omega\right)\right|^{2}}\\
\times e^{+i\intop_{0}^{t}d\tau s_{-}^{*}\left(\tau\right)z_{\textrm{emit}}\left(\tau\right)-i\intop_{0}^{t}d\tau s_{+}\left(\tau\right)z_{\textrm{emit}}^{*}\left(t\right)},\label{eq:emissive_stochastic_unravelling}
\end{multline}
where the following $c$-number stochastic field was introduced
\begin{equation}
z_{\textrm{emit}}\left(\tau\right)=\intop_{-\infty}^{+\infty}d\omega c\left(\omega\right)e^{-i\omega\tau}\xi\left(\omega\right),
\end{equation}
and $\xi\left(\omega\right)$ is a complex white noise,
\begin{equation}
\overline{\xi\left(\omega\right)\xi^{*}\left(\omega^{\prime}\right)}=\delta\left(\omega-\omega^{\prime}\right).
\end{equation}
For the excitation part of influence functional, we have: 
\begin{multline}
e^{-\iintop_{0}^{t}d\tau d\tau^{\prime}\boldsymbol{s}^{\dagger}\left(\tau\right)\mathbf{K}_{\textrm{exc}}\left(\tau-\tau^{\prime}\right)\boldsymbol{s}\left(\tau^{\prime}\right)}\\
=e^{-\iintop_{0}^{t}d\tau d\tau^{\prime}\left\{ s_{+}^{*}\left(\tau\right)-s_{-}^{*}\left(\tau\right)\right\} M_{\textrm{exc}}\left(\tau-\tau^{\prime}\right)\left\{ s_{+}\left(\tau\right)-s_{-}\left(\tau\right)\right\} }\\
=\int D\left[\eta\right]e^{-\intop_{-\infty}^{+\infty}d\omega\left|\eta\left(\omega\right)\right|^{2}}\\
\times e^{-i\intop_{0}^{t}d\tau\left\{ s_{+}^{*}\left(\tau\right)v_{\textrm{exc}}\left(\tau\right)+s_{+}\left(\tau\right)v_{\textrm{exc}}^{*}\left(\tau\right)\right\} }\\
\times e^{+i\intop_{0}^{t}d\tau\left\{ s_{-}^{*}\left(\tau\right)v_{\textrm{exc}}\left(\tau\right)+s_{-}\left(\tau\right)v_{\textrm{exc}}^{*}\left(\tau\right)\right\} },\label{eq:thermal_stochastic_unravelling}
\end{multline}
 where the following $c$-number stochastic field was introduced
\begin{equation}
v_{\textrm{exc}}\left(\tau\right)=\intop_{-\infty}^{+\infty}d\omega\sqrt{n\left(\omega\right)}c\left(\omega\right)e^{-i\omega\tau}\eta\left(\omega\right),
\end{equation}
and $\eta\left(\omega\right)$ is another (independent from $\xi\left(\omega\right)$)
complex white noise,
\begin{equation}
\overline{\eta\left(\omega\right)\eta^{*}\left(\omega^{\prime}\right)}=\delta\left(\omega-\omega^{\prime}\right).
\end{equation}
Now, if we substitute the stochastically unraveled parts Eq. (\ref{eq:emissive_stochastic_unravelling})
and (\ref{eq:thermal_stochastic_unravelling}) into the expression
for the mean value of the impurity observable Eq. (\ref{eq:open_system_keldysh_integral}),
we obtain
\begin{multline}
\left\langle \widehat{O}\left(t\right)\right\rangle \\
=\overline{\int\left[D\left[q_{+},q_{-}\right]e^{iS_{\textrm{stoch}}\left[q_{+},q_{-}\right]}O\left(q_{+}\left(t\right),q_{-}\left(t\right)\right)\right]}_{\xi,\eta},\label{eq:open_system_keldysh_integral_unraveled}
\end{multline}
where
\begin{multline}
S_{\textrm{stoch}}\left[q_{+},q_{-}\right]=S_{\textrm{sys}}\left[q_{+},q_{-}\right]\\
-\intop_{0}^{t}d\tau\left\{ s_{+}^{*}\left(\tau\right)v_{\textrm{exc}}\left(\tau\right)+s_{+}\left(\tau\right)\left(v_{\textrm{exc}}^{*}\left(\tau\right)+z_{\textrm{emit}}^{*}\left(t\right)\right)\right\} \\
+\intop_{0}^{t}d\tau\left\{ s_{-}^{*}\left(\tau\right)\left(v_{\textrm{exc}}\left(\tau\right)+z_{\textrm{emit}}\left(t\right)\right)+s_{-}\left(\tau\right)v_{\textrm{exc}}^{*}\left(\tau\right)\right\} \\
+i\iintop_{0}^{t}d\tau d\tau^{\prime}\boldsymbol{s}^{\dagger}\left(\tau\right)\mathbf{K}_{\textrm{virt}}\left(\tau-\tau^{\prime}\right)\boldsymbol{s}\left(\tau^{\prime}\right).\label{eq:stochastic_action}
\end{multline}
Now we are almost done. In order to find the numerical algorithm which
follows from Eqs. (\ref{eq:open_system_keldysh_integral_unraveled})-(\ref{eq:stochastic_action}),
we switch back to the operator representation. This is done by finding
the Hamiltonian interpretation of the action functional Eq. (\ref{eq:stochastic_action}).
The first three lines of Eq. (\ref{eq:stochastic_action}) are interpreted
as the Keldysh-contour evolution under the non-Hermitian Hamiltonian
\begin{equation}
\widehat{H}_{\textrm{sys}}+\hat{s}\left\{ z_{\textrm{emit}}^{*}\left(\tau\right)+v_{\textrm{exc}}^{*}\left(\tau\right)\right\} +\hat{s}^{\dagger}v_{\textrm{exc}}\left(\tau\right).
\end{equation}
In order to interpret the fourth line of Eq. (\ref{eq:stochastic_action}),
we remember the discussion at the end of section \ref{subsec:Influence-functional-and}
that the influence functional with $\mathbf{K}_{\textrm{virt}}\left(\tau-\tau^{\prime}\right)$
corresponds to the full impurity-bath quantum problem Eq. (\ref{eq:in-out-problem})
with the vacuum-vacuum boundary conditions for the bath, Eqs. (\ref{eq:vacuum_input_state})-(\ref{eq:vacuum_output_state}).
Therefore, the stochastically unraveled average Eq. (\ref{eq:open_system_keldysh_integral_unraveled})
can be written in the operator language as
\begin{equation}
\left\langle \widehat{O}\left(t\right)\right\rangle =\overline{\left\langle \psi_{\textrm{dress}}\left(t\right)\right|\left.0_{\textrm{b}}\right\rangle \widehat{O}\left\langle 0_{\textrm{b}}\right.\left|\psi_{\textrm{dress}}\left(t\right)\right\rangle }_{\xi,\eta},
\end{equation}
where $\psi_{\textrm{dress}}\left(t\right)$ is the impurity wavefunction
``dressed'' by virtual excitations
\begin{equation}
\left|\psi_{\textrm{dress}}\left(t\right)\right\rangle ==\mathcal{T}e^{-i\intop_{0}^{t}\widehat{H}_{\textrm{stoch}}\left(\tau\right)d\tau}\left|0_{\textrm{bath}}\right\rangle \otimes\left|\psi_{\textrm{sys}}\left(0\right)\right\rangle ,
\end{equation}
here $\mathcal{T}$ is the usual time ordering, and the stochastic
Hamiltonian $\widehat{H}_{\textrm{stoch}}\left(\tau\right)$ is 
\begin{multline}
\widehat{H}_{\textrm{stoch}}\left(\tau\right)=\widehat{H}_{\textrm{sys}}+\hat{s}\left\{ \widehat{b}^{\dagger}+z_{\textrm{emit}}^{*}\left(\tau\right)+v_{\textrm{exc}}^{*}\left(\tau\right)\right\} \\
+\hat{s}^{\dagger}\left\{ \widehat{b}+v_{\textrm{exc}}\left(\tau\right)\right\} .
\end{multline}
The dressed wavefunction $\psi_{\textrm{dress}}\left(t\right)$ is
the solution of the non-Markovian stochastic Schrodinger equation:
\begin{equation}
\partial_{t}\psi_{\textrm{dress}}\left(t\right)=-i\widehat{H}_{\textrm{stoch}}\left(\tau\right)\psi_{\textrm{dress}}\left(t\right)\label{eq:NMSSE}
\end{equation}
with initial conditions 
\begin{equation}
\left|\psi_{\textrm{dress}}\left(0\right)\right\rangle =\left|0_{\textrm{b}}\right\rangle \otimes\left|\psi_{\textrm{s}}\left(0\right)\right\rangle .
\end{equation}
Observe that formally we still have the full quantum problem for the
bath. However, since most of the quantum entanglement is eliminated
by performing the averaging over the classical noises $\xi$ and $\eta$,
and only the projection to the bath vacuum is required, we expect
much faster convergence when applying numerical discretizations to
Eq. (\ref{eq:NMSSE}). 

\subsection{\label{subsec:Approximate-solution-of}Numerical solution of the
stochastic dressed wavefunction equation}

The dressed wavefucntion $\psi_{\textrm{dress}}\left(t\right)$ is
calculated in a truncated Fock space by keeping all the relevant states
of the impurity and all the bath states with at most $N$ excitations,
for a certain fixed $N$. Note that when a truncation of the Hilbert
space is applied, the norm of the reduced impurity density matrix
is not conserved:
\begin{equation}
Z\left(t\right)=\textrm{Tr}_{\textrm{i}}\widehat{\rho}_{\textrm{i}}=\overline{\left\Vert \left\langle 0_{\textrm{b}}\right.\left|\psi_{\textrm{dress}}\left(t\right)\right\rangle \right\Vert ^{2}}_{\xi,\eta}\neq1.
\end{equation}
We compensate for this by normalizing the computed observable averages:
\begin{equation}
\left\langle \widehat{O}\left(t\right)\right\rangle =\frac{\overline{\left\langle \psi_{\textrm{dress}}\left(t\right)\right|\left.0_{\textrm{b}}\right\rangle \widehat{O}\left\langle 0_{\textrm{b}}\right.\left|\psi_{\textrm{dress}}\left(t\right)\right\rangle }_{\xi,\eta}}{\overline{\left\Vert \left\langle 0_{\textrm{b}}\right.\left|\psi_{\textrm{dress}}\left(t\right)\right\rangle \right\Vert ^{2}}_{\xi,\eta}}.
\end{equation}

\section{\label{sec:RESULTS}RESULTS}

\begin{figure*}
\includegraphics[width=2\columnwidth]{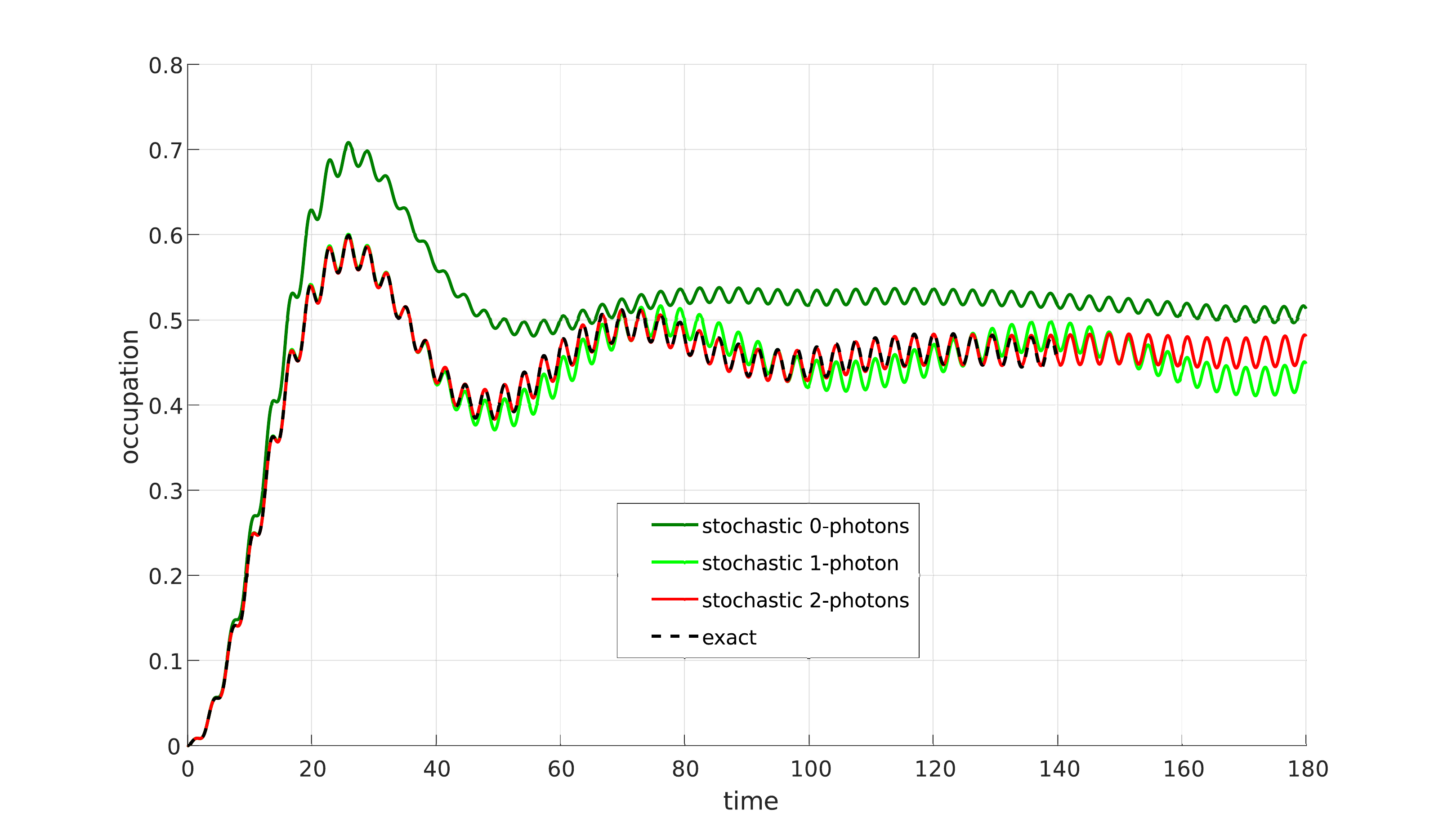}

\caption{\label{fig:figure_convergence_band_centered} The impurity energy
level is at the center of bath's energy band.The exact mean occupation
of the equivalent qubit (dashed black line) is computed in the truncated
Fock space, which required to take into account 8 bath excitations
to converge up to the time $t=180$. At the same time, the results
for stochastic dressed wavefunction method at $N=0$ (dark olive green),
$N=1$ (light green), and $N=2$ (red), show that we reach convergence
up to the stationary regime with only 2 virtual excitations.}

\end{figure*}

We test the proposed stochastic dressed wavefunction approach on the
driven spin-boson model,
\begin{equation}
\widehat{H}_{\textrm{sys}}=\frac{\varepsilon}{2}\widehat{\sigma}_{z}+\widehat{\sigma}_{+}f\left(t\right)+\widehat{\sigma}_{-}f^{*}\left(t\right),
\end{equation}
coupled through the spin impurity operator 
\begin{equation}
\widehat{s}=\widehat{\sigma}_{-},
\end{equation}
to the bath with the semicircle density of states
\begin{equation}
c\left(\omega\right)=\theta\left(\left|\omega-\varepsilon_{0}\right|-2h\right)\sqrt{\frac{1}{4\pi}\left\{ 4h^{2}-\left(\omega-\varepsilon_{0}\right)^{2}\right\} },
\end{equation}
which corresponds to a chain of bose sites with on-site energy $\varepsilon_{0}$
and hopping between the sites $h$. For calculations, we use the following
values of parameters of the bath: $\varepsilon_{0}=1$, $h=0.05$.
The driving field is defined as
\begin{equation}
f\left(t\right)=0.1\cos t.
\end{equation}
We consider the two cases, with the bath initially at zero temperature.
The first case is when the impurity energy level is placed at the
center of the bath's energy band:
\begin{equation}
\varepsilon=\varepsilon_{0}=1.
\end{equation}
We calculated the occupation 
\begin{equation}
\widehat{n}=\widehat{\sigma}_{+}\widehat{\sigma}_{-}
\end{equation}
of the equivalent qubit. In Fig. \ref{fig:figure_convergence_band_centered},
we present the convergence of stochastic dressed wavefunction results
with $N=0$ (only impurity Hilbert space, no virtual excitations),
$N=1$, and $N=2$, to the exact results in the truncated Fock space. 

\begin{figure*}
\includegraphics[width=2\columnwidth]{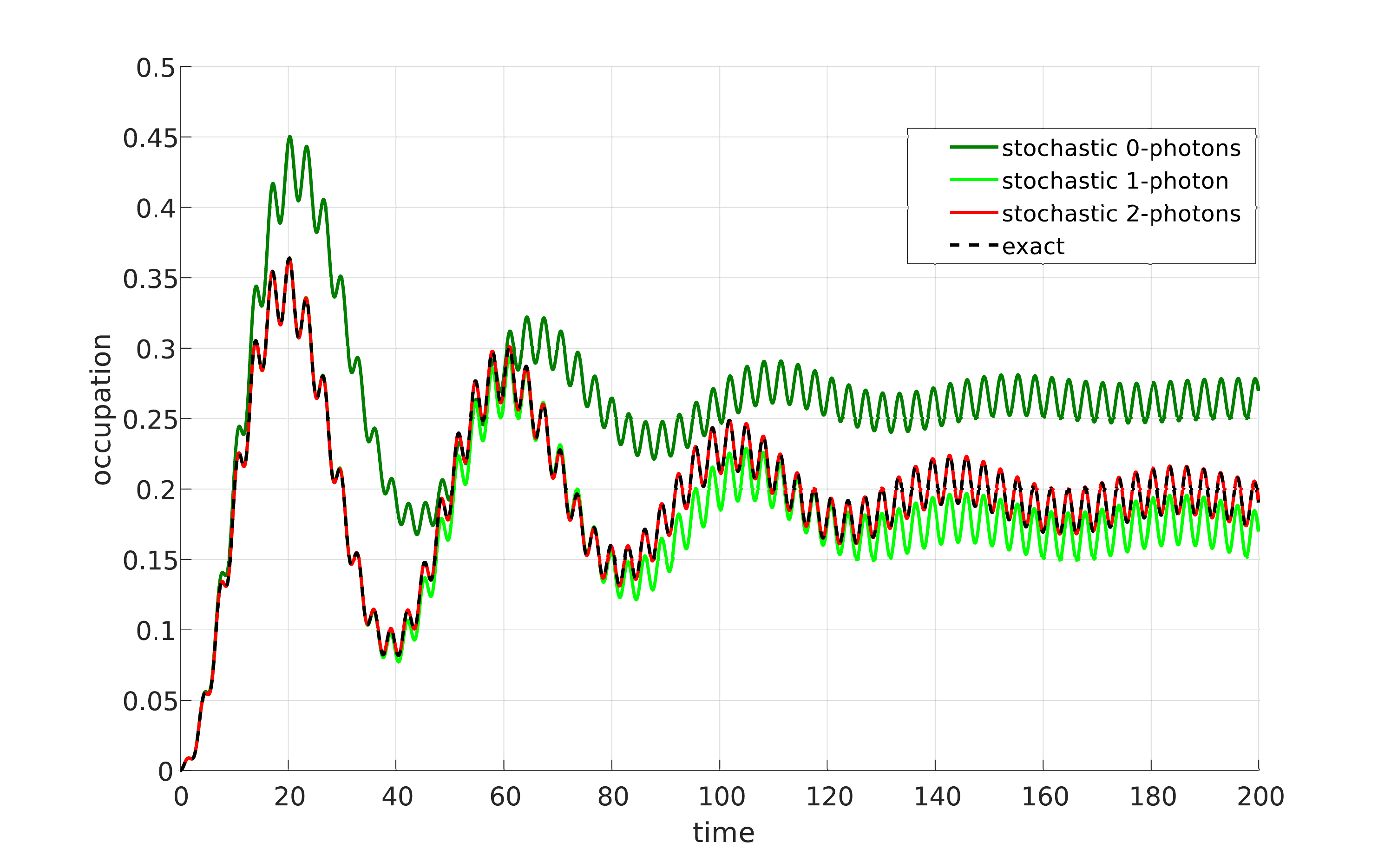}

\caption{\label{fig:figure_convergence_edge_of_band}The impurity energy level
is at the edge of bath's energy band. Lines and colors have the same
meaning as in previous figure. The exact result required 8 bath excitations,
whereas the stochastic dressed wavefunction converge uniformly with
only 2 virtual excitations.}

\end{figure*}

The second case we considered is when the impurity energy level is
placed at the edge of the bath energy band:
\begin{equation}
\varepsilon=\varepsilon_{0}-2h.
\end{equation}
In Fig. \ref{fig:figure_convergence_edge_of_band} we show the convergence
of results for the occupation of the equivalent qubit. In both cases
the virtual excitations in $\psi_{\textrm{dress}}\left(t\right)$
were taken into account by including the first 20 sites of the bozonic
chain. 

From the presented results we see that the convergence on the whole
time interval is achieved with only two virtual excitations, whereas
ED required to include the states with 8 excitations of the bath.
This result confirms our idea that the stochastic dressed wavefunction
method is capable of alleviating the exponential complexity of the
real-time simulation. 

Our approach is related to the conventional non-Markovian quantum
state diffusion (NMQSD) methods \cite{Diosi1997,Diosi1998,Strunz2001,Hartman2017},
but there is important difference between them. NMQSD includes only
the impurity degrees of freedom, and the influence of the virtual
cloud is represented through the functional derivative of the stochastic
trajectory with respect to the noise. Since the functional derivative
is a computationally complex object, a hierarchy of approximations
is developed \cite{Hartman2017}. However, it is difficult to judge
apriori how fast such a hierarchy would converge in the strong coupling
regime. At the same time, the stochastic dressed wavefunction method
takes into account the fact that the physical state of any open system
is not restricted to the open system's degrees of freedom, but surrounded
by a cloud of virtual excitations. This way we obtain a clear physical
picture of the major convergence factor: the dimension of the part
of virtual cloud which is entangled to the impurity and which is statistically
significant.

\section{\label{sec:CONCULSION}CONCULSION}

In this work we present a novel numerically-exact simulation approach
for the dynamics of quantum impurity models: the stochastic dressed
wavefunction method. In this method, all the observable effects of
the environment (irreversibly emitted excitations and the excitations
due to finite occupation of the bath modes) are calculated by a Monte
Carlo procedure without the sign problem. At the same time the unobservable
virtual excitations are calculated by an exact diagonalization. We
illustrate our method by providing the results of test calculations
for the driven spin-boson model: only two virtual excitations are
enough to achieve the uniform convergence on a large time interval.
\begin{acknowledgments}
The study was founded by the RSF, grant 16-42-01057.
\end{acknowledgments}

\bibliographystyle{apsrev4-1}

\end{document}